\documentclass[fleqn,10pt]{wlscirep}
\usepackage[utf8]{inputenc}
\usepackage[T1]{fontenc}
\usepackage{syntax}
\usepackage{graphicx}
\usepackage{amsmath}
\usepackage{amssymb}
\usepackage{tcolorbox}

\title{Coronavirus Covid--19 spreading in Italy: optimizing an epidemiological model with dynamic social distancing through Differential Evolution}

\author[1.*]{I. De Falco}
\author[2,1]{A. Della Cioppa}
\author[1]{U. Scafuri}
\author[1]{E. Tarantino}
\affil[1]{National Research Council of Italy (CNR) \\ Institute for High--Performance Computing and Networking (ICAR) \\ Via Pietro Castellino 111, 80131 Naples, Italy}
\affil[2]{Department of Information Engineering, Electrical Engineering and Applied Mathematics\\
University of Salerno
\\
Via Giovanni Paolo II, 132, 84084 Fisciano (SA), Italy}

\affil[*]{ivanoe.defalco@icar.cnr.it}

\date{January 2020}

\begin{abstract}
The aim of this paper consists in the application of a recent epidemiological model, namely SEIR with Social Distancing (SEIR--SD), extended here through the definition of a social distancing function varying over time, to assess the situation related to the spreading of the coronavirus Covid--19 in Italy and in two of its most important regions, i.e., Lombardy and Campania. To profitably use this model, the most suitable values of its parameters must be found. The estimation of the SEIR--SD model parameters takes place here through the use of Differential Evolution, a heuristic optimization technique. In this way, we are able to evaluate for each of the three above--mentioned scenarios the daily number of infectious cases from today until the end of virus spreading, the day(s) in which this number will be at its highest peak, and the day in which the infected cases will become very close to zero.
\end{abstract}


\begin{document}

\maketitle

\section{Introduction}

Several compartmental models have been designed in epidemiology to simplify the mathematical modeling of infectious disease, so as to describe the spreading of diseases in a population of individuals. Among them, we can recall here at least the SIR \cite{kermack1927contribution} and the SEIR \cite{diekmann2012mathematical}, where SIR is the basic such model, whereas SEIR expands it. In these days, an extension to SEIR has been proposed by Hubbs, called SEIR with Social Distancing (SEIR-SD) \cite{hubbs2020social}. As the name suggests, this latter model also accounts for issues related to keeping individuals as distanced as possible one from another. 

In this paper we make use of SEIR-SD and we extend it through the definition of a social distancing function varying over time. To have this model working properly in a way that it can describe a real--world situation, the exact values of the parameters must be found. In fact, as it can be imagined, even slight differences in these values will lead to a completely different evolution of the disease spread described by the equations. 

Finding this set of values is a far--from--easy task, as a huge number of possible such combinations exists. Technically speaking, this is an $NP$--hard problem, and is the main limitation to the practical use of the above described models.

Yet, with the advances of the research, this task can be suitably faced through the use of heuristic optimization techniques.

From among the many available such techniques, we have decided here to avail ourselves of Differential Evolution (DE). Its description is beyond the scope of this paper, and the interested reader can find useful information in 
\cite{price2006differential}. Suffice it to say here that DE is a Machine Learning technique widely used in Artificial Intelligence. It is a heuristic population--based optimization technique that has proved itself extremely useful to find optimal or sub--optimal solutions to multidimensional real-valued problems, even multi--objective and constrained, taken from many different fields. This means that DE can often find a good set of values for the variables of these problems, even for large sets up to, say, hundreds of variables,

We wish to apply these models the spreading of coronavirus Covid--19 that is taking place in these days in Italy, consequently we have downloaded the related data, that is publicly available at \cite{dati-regioni} thanks to Italian Ministry for Health.

Our goal is to be able to evaluate the daily number of infectious cases from today until the end of virus spreading, the day(s) in which this number will be at its highest peak, and the day in which the infected cases will become very close to zero.

We will also investigate the consequences of variations in the social distancing level on the evolution over time of Covid--19 spreading in Italy. This study could serve as a useful guideline to Italian Government as well to the Government of any other State in which Covid--19  spreading is occurring.

\section{The models}

\subsection{The SIR model}

The basic SIR model is composed by three compartments, and takes into account at any given time $t$ the number of susceptible individuals ($S$), the number of infectious individuals ($I$), and the number of recovered individuals (R). Hence it is based on three time functions $S(t)$, $I(t)$ and $R(t)$. These functions are normalized so that at any given time $t$ the following holds: $S(t) + I(t) + R(t) =1$.

This model is reasonably predictive for infectious diseases which are transmitted from human to human, and where recovery confers lasting resistance.

It can be represented by the following set of differential equations:

\begin{align}
\frac{dS}{dt} & = \frac{-\beta \cdot I \cdot S}{N} \\[5pt]
\frac{dI}{dt} & = \frac{\beta \cdot I \cdot S}{N} - \gamma \cdot I \\[5pt]
\frac{dR}{dt} & = \gamma \cdot I \\
\end{align}

\noindent where $N$ is the sum of $S$, $I$ and $R$. The non--negative real--valued parameter $\beta$  is the average contact rate in the population, whereas the non--negative real--valued parameter $\gamma$ is the inverse of the mean infectious period ($1/t_{infectious}$). If we impose that the sum of the three equations above must be equal to 0, we obtain that the population size $N$ remains constant,

As it can be seen, this model contains two parameters, i.e., $\beta$ and $\gamma$.

It should be remarked that in this model, as well as in all the other ones described in the following, $R$ represents the sum both of the individuals who actually return healthy after being infected and of those who die due to the epidemic.

A very important parameter in SIR model and in its derivations is the ratio $R_0 = \frac{\beta}{\gamma}$, called basic reproduction ratio. It represents the expected number of currently susceptible individuals that will be infected by an infectious one, so it accounts for the degree of infectiousness of the specific epidemic being examined.

\subsection{The SEIR model}

This model represents an extension to the SIR model. In fact, it takes into account the fact that there exists an incubation period in which the individual has been infected but not yet infectious. This is represented by a further compartment containing such exposed ($E$)individuals.
SEIR model also accounts for vital dynamics in terms of birth rate and death rate, so that the population size can vary. Actually, given the short lifetime considered here, we are not interested in this feature, so that $S + E + I + R = N$, and the equations of this model reduce to:

\begin{align}
\frac{dS}{dt} & = \frac{-\beta \cdot I \cdot S}{N} \\[5pt]
\frac{dE}{dt} & = \frac{\beta \cdot I \cdot S}{N} - \alpha \cdot E \\[5pt]
\frac{dI}{dt} & = \alpha \cdot E - \gamma \cdot I \\[5pt]
\frac{dR}{dt} & = \gamma \cdot I
\end{align}

\noindent where the non--negative real--valued parameter $\alpha$ is the inverse of the incubation period ($1/t_{incubation}$).
Hence, the dynamics of this model is completely determined by setting the values for the thre parameters $\alpha$, $\beta$ and $\gamma$.

\subsection{The SEIR Model vith Social Distancing}

With the aim at reducing the spreading of Covid--19, as in many other countries in which Covid--19 virus is present, also in Italy the Government has decided to enforce {\it social distancing}. Basically, many shops considered unnecessary have been closed, and so have been schools and universities. Moreover, smart working has been allowed wherever possible. Furthermore, in all public spaces left open to people, as food shops and supermarkets and pharmacies, and in the streets as well, a distance of at least one meter should be kept between people.

The way this influences the spreading of the virus has been modeled in several ways. We will make reference here to the approach proposed in these days by Hubbs \cite{hubbs2020social}, and will shortly refer to it as SEIR with Social Distancing (SEIR-SD). This is based on the introduction of a new real--valued parameter $\rho$, ranging within 0.0 and 1.0.The value 0.0 represents the ideal case in which everyone is locked down in quarantine,  whereas the value 1.0 reduces this model to the SEIR model in which no social distancing is considered. The introduction of $\rho$ implies the modification in the SEIR model of the equations related to $S$ and $E$, so that the SEIR-SD model is the following:

\begin{align}
\frac{dS}{dt} & = - \frac{\rho \cdot \beta \cdot S \cdot I}{N} \\[5pt]
\frac{dE}{dt} & = \frac{\rho \cdot \beta \cdot S \cdot I}{N} - \alpha \cdot E \\[5pt]
\frac{dI}{dt} & = \alpha \cdot E - \gamma \cdot I \\[5pt]
\frac{dR}{dt} & = \gamma \cdot I
\end{align}

Strictly speaking, $\rho$ is another parameter of the model, so that this depends on four parameters, i.e., $\alpha$, $\beta$, $\gamma$ and $\rho$. Their values should be suitably found, so that the model can closely represent the situation being studied.

\subsection{SEIR--SD Model with Dynamic Social Distancing}

We have noticed that keeping a value of $\rho$ constant during the whole evolution of the pandemic over time is unrealistic.
Rather, what happened in the Italian case is that in the first days, in which we are aware of just few cases, no social distancing was enforced. Then, after a few days, some general rules of thumb were suggested, as avoiding unnecessary travels and exits from homes and, in such cases, keeping at a distance of at least one metre from other people. As the situation started to get worse, some parts of Italy were isolated, and later on movements within Italy were more and more limited. With time, more and more activities and shops were closed by Government. Finally, even going out from homes was very strictly limited by law. Of course, this cannot be represented by a single value of $\rho$ kept constant over time.

\begin{figure}[t!]
	\centering
	\includegraphics[width=10cm]{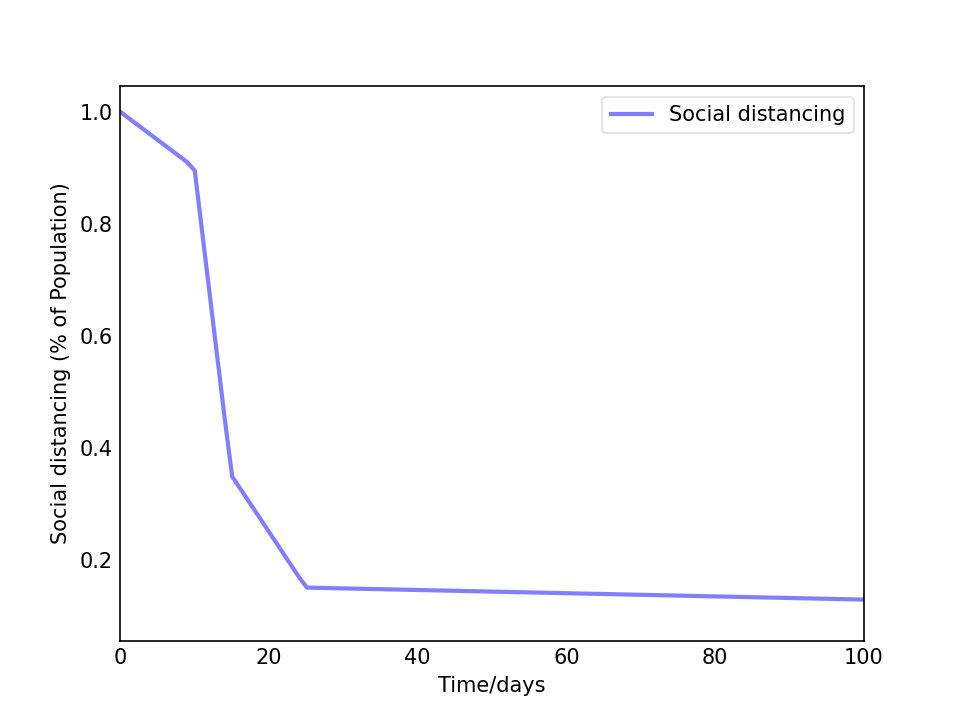}
	\caption{Time-varying function used to represent dynamic social distancing.}
	\label{function}
\end{figure}

We have designed a time-varying social distancing function, shown in Figure \ref{function}, by considering the different decrees issued on different dated by Italian First Minister, which have led over time to lower and lower freedom of movement for Italian citizens. Then, a univariate spline has been used to obtain a 1-dimensional smoothing fitting.  

Such a function has been used to obtain a used for the 1-dimensional smoothing spline fitting can be observed in Figure \ref{function}.

%
%

In the following of our experiments, we will always make use of this function.

\section{The Simulations}

\subsection{The Methodology}

The simulations are carried out by using the SEIR--SD model on the set of infection cases taking place in Italy in these days, as they can be downloaded from the Italian Ministry for Health free repository \cite{dati-regioni}, updated to March 29. 2020.

From an optimization point of view, we have to find the best possible set of values for the SEIR-SD model parameters that allows us to follow as closely as possible the series of the real infection cases taking place in Italy in these days.
As already said above, we make use here of Differential Evolution. 

Given a real--valued optimization problem with size $L$, Differential Evolution is an optimization method that starts by randomly creating an initial set of possible solutions to the problem, each represented as a vector of $L$ real--valued numbers. 
The cardinality of the population is termed {\it population size} ($P_s$) and is kept constant during the evolution. 
Each time a solution is created, its quality at solving the problem is evaluated by means of a suitable fitness function $\Phi$, that must be optimized.
Then starting from the current population, a new one is created for the next generation thanks to the use of a suitable mutation mechanism. 
Many different mechanisms exist, basically each of them starts from the generic i--th individual, receives in input a set of other randomly--chosen current individuals (two, three, or four) and, based on the values of two parameters, named {\it mutation factor} ($F$) and {\it crossover ratio} ($C_r$), mixes their features to those of the i--th individual so as to obtain a new trial individual. 
This latter is compared to the corrent i--th one, and the better enters the new population under construction.
This mechanism is repeated for $P_s$ times at each generation, so that at the end we will obtain the new $P_s$--sized  new population for next generation. 
This is repeated for a number of generations represented by $G$. 
The individual with the best fitness value in the final population is the solution found by the algorithm.


For the problem at hand, Differential Evolution should find the most suitable values for the three parameters $\alpha$, $\beta$ and $\gamma$. We find these values by taking as the goal of the optimization process the minimization of the Root Mean Square Error (RMSE) between the number of infectious cases estimated by the SEIR-SD model $I_e$ and the actual number of infectious cases in Italy $I_a$, where this RMSE is computed over the number of days $N_d$ starting from the official onset of Covid--19 in Italy and the last day for which we have actual data available. In formulae:

\begin{equation}
RMSE = \sqrt{\frac{1}{N_d} \cdot \sum_{i=1}^{N_d} (I_a - I_e)^2}
\end{equation}

Once obtained in this way the set of parameter values minimizing the RMSE, by running the SEIR-SD with those values we will obtain the evolution of Covid--19 in Italy. Consequently, we will be able to evaluate the daily number of infectious cases from today until the end of virus spreading. Interestingly, we will be able to estimate the day(s) in which this number will be at its highest peak, as well as the day in which the infected cases will become very close to zero.

Based on our experience in the use of Differential Evolution, and after a short preliminary phase of parameter tuning, we have decided to use the following values for the parameters of the Differential Evolution optimizer: population size 50, number of generations 10,000, rand/1/bin as the mutation mechanism, mutation probability 0.7, and recombination probability 0.9.


\subsection{The Numerical Results}

\subsubsection{Modeling Italy} 

Italian population as of end 2019 is equal to 60,359,546 people. 

The execution of Differential evolution has allowed us to determine the most suitable values for the SEIR--SD model as follows: $\alpha = 0.33$, $\beta = 2.11$ and $\gamma = 0.50$. These values correspond to a value of the RMSE equal to $9.12^{-10}$.
The value for $R_0$ is equal to 4.2.

\begin{figure}[t!]
	\centering
	\includegraphics[width=10cm]{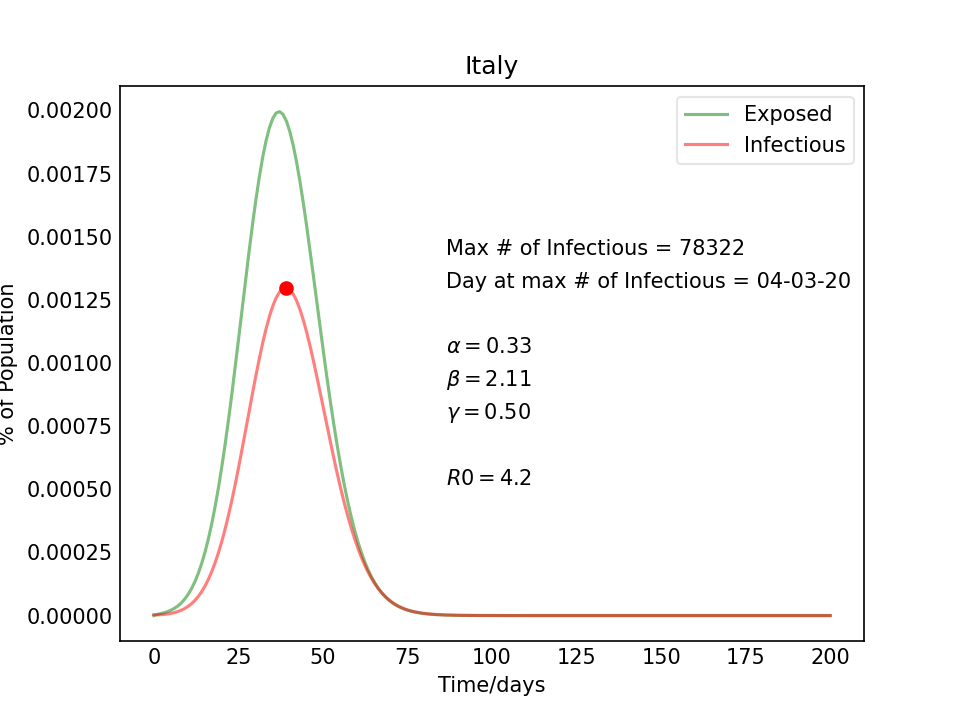}
	\caption{Daily infectious cases: the data computed by our approach and the data for Italy.}
	\label{figure-italy-1}
\end{figure}

As it can be seen in Fig. \ref{figure-italy-1}, the data provided by the model are in good accordance with the real--world data.

\begin{figure}[t!]
\centering
\includegraphics[width=10 cm]{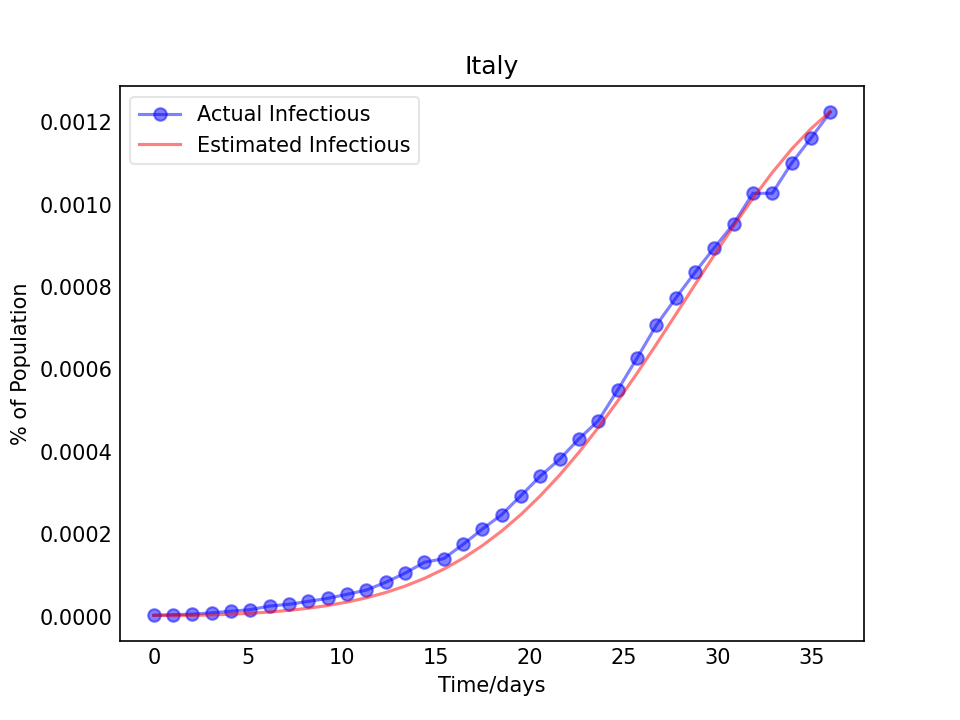}
\caption{The evolution over time of Covid--19 epidemic in Italy based on the parameter values found for the SEIR--SD model by Differential Evolution.}
\label{figure-italy-2}
\end{figure}

The use of these values leads to model the diffusion of coronavirus Covid--19 in Italy as it is shown in Fig. \ref{figure-italy-2}.

From the figure we can appreciate that the peak of the infection is expected to take place around April 3, with a number of infectious subjects equal to about 78.322. Then, this number tends to decrease, and approaches zero around June 17.


\subsubsection{Modeling Lombardy}

We have run the same experiments by taking into account Lombardy region. This is important because it is the most  populated Italian region with about 10 million inhabitants (10,060,574 is the official data for 2019). Moreover, Lombardy registered the onset in Italy, has currently the highest number of infected subjects, and registers the highest number of deaths due to this epidemics. 

Following the general procedure shown above, we have run the Differential Evolution algorithm tofind the best parameter values for the SEIR-SD mode.
%
For region Lombardy the set of parameters has resulted to be: $\alpha = 0.33$, $\beta=1.96$ and $\gamma=0.50$, with an RMSE value equal to $7.9^{-9}$. The value for $R_0$ is equal to 3.9.

\begin{figure}[t!]
	\centering
	\includegraphics[width=10 cm]{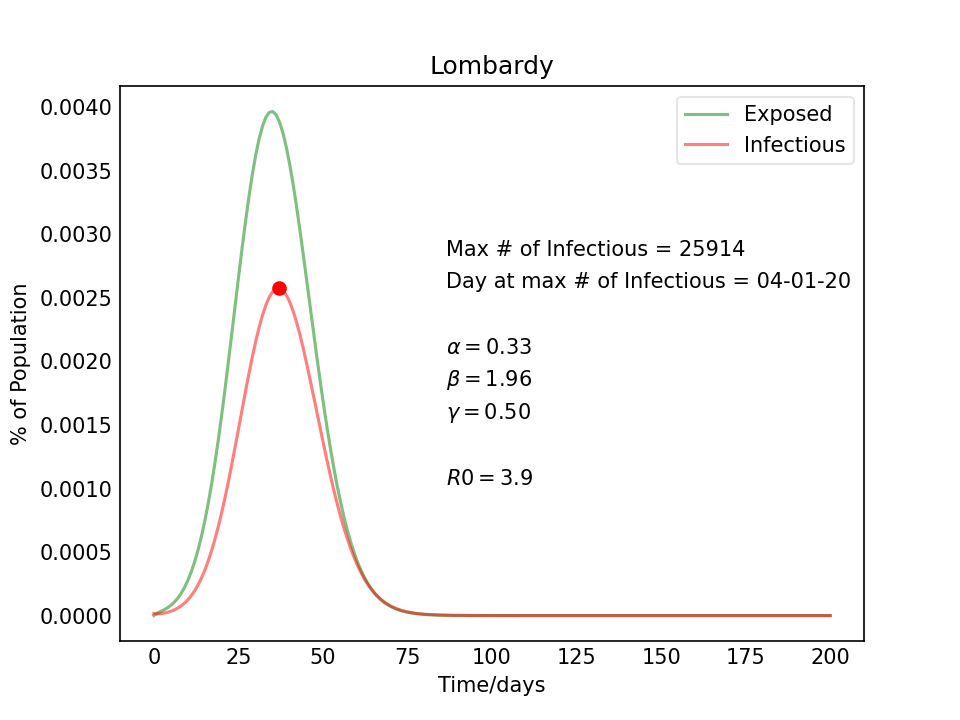}
	\caption{Daily infectious cases: the data computed by our approach and the data for Lombardy.}
	\label{figure-lombardy-1}
\end{figure}

As it can be seen in Fig. \ref{figure-lombardy-1}, the data provided by the model are in good accordance with the real--world data.

\begin{figure}[t!]
	\centering
	\includegraphics[width=10 cm]{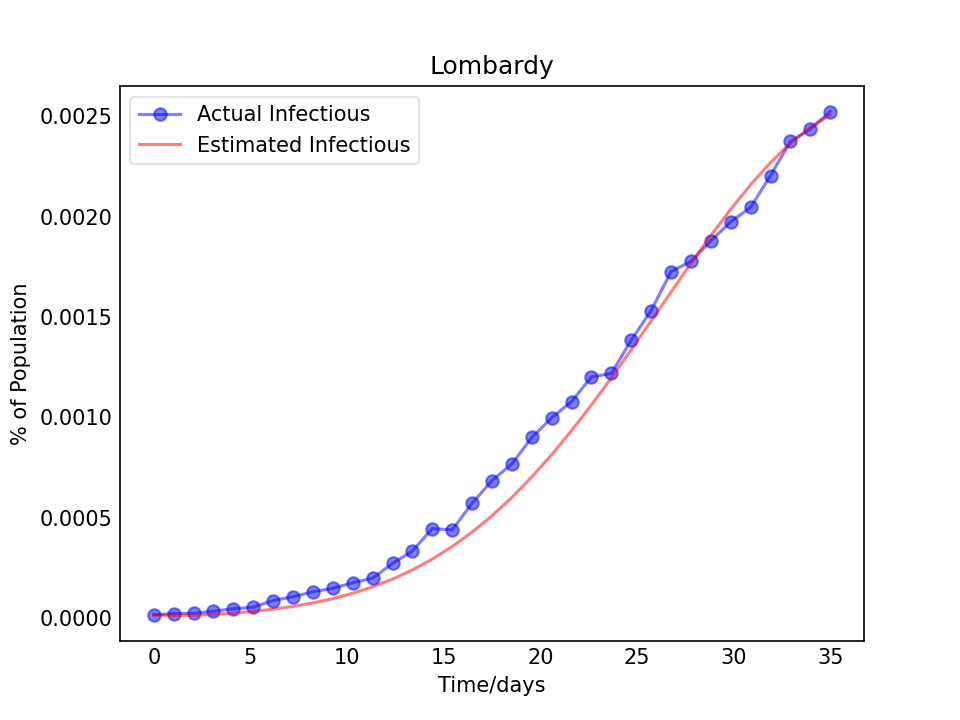}
	\caption{The evolution over time of Covid--19 epidemic in Lombardia based on the parameter values found for the SEIR--SD model by Differential Evolution.}
	\label{figure-lombardy-2}
\end{figure}

The use of these values leads to model the diffusion of coronavirus Covid--19 in Lombardy as it is shown in Fig. \ref{figure-lombardy-2}.


From the figure we can appreciate that the peak of the infection is expected to take place in Lombardy around April 1, with a number of infectious subjects equal to about 25.914. Then, thin number tends to decrease, and approaches zero around June 7.

\subsubsection{Modeling Campania}

The investigation of Campania is important because Campania is the third most populated region in Italy, with almost six million inhabitants (5,861,529 is the official data for 2019). Moreover, the interest lies in the fact that Campania has been infected much later than Lombardy, is quite distant from this latter, and as of today shows a much lower number of infectious cases. Therefore, we wonder whether the epidemics evolution in Campania can be different from that in Lombardy.

Also in this case we have run the Differential Evolution algorithm to find the best parameter values for the SEIR-SD model by using the data for the infectious subjects in Campania.
%
By running our experiments on the data for Campania region, instead, the set of parameters has resulted to be: $\alpha = 0.33$, $\beta=2.25$ and $\gamma=0.50$, with an RMSE value equal to $2.18^{-10}$. The value for $R_0$ is equal to 4.5.

\begin{figure}[t!]
	\centering
	\includegraphics[width=10 cm]{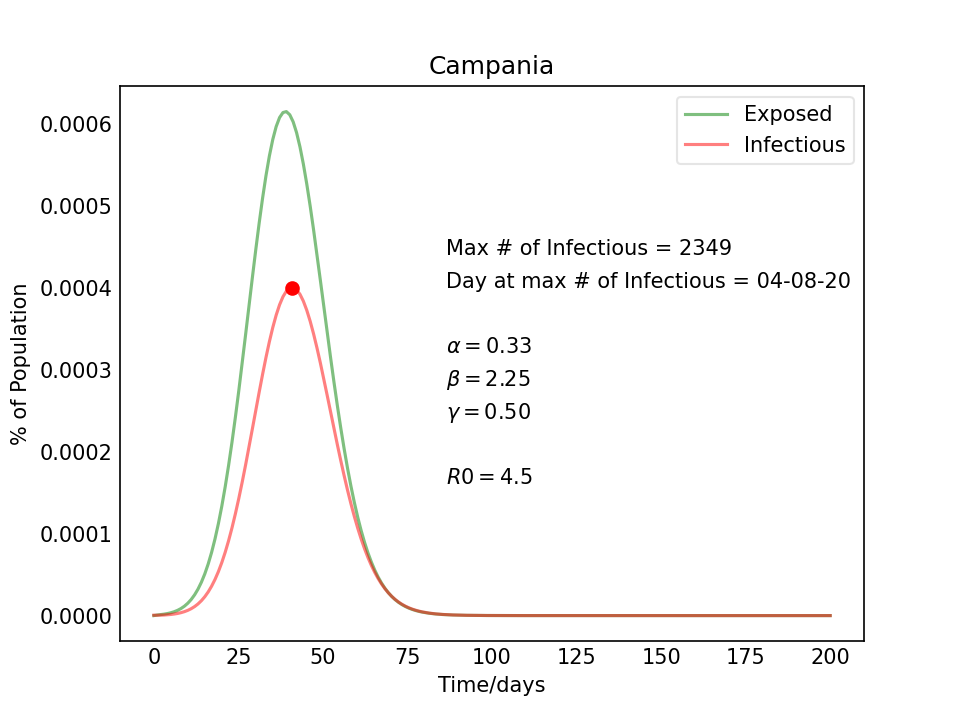}
	\caption{Daily infectious cases: the data computed by our approach and the data for Campania.}
	\label{figure-campania-1}
\end{figure}

As it can be seen in Fig. \ref{figure-campania-1}, the data provided by the model are in good accordance with the real--world data.

\begin{figure}[t!]
	\centering
	\includegraphics[width=10 cm]{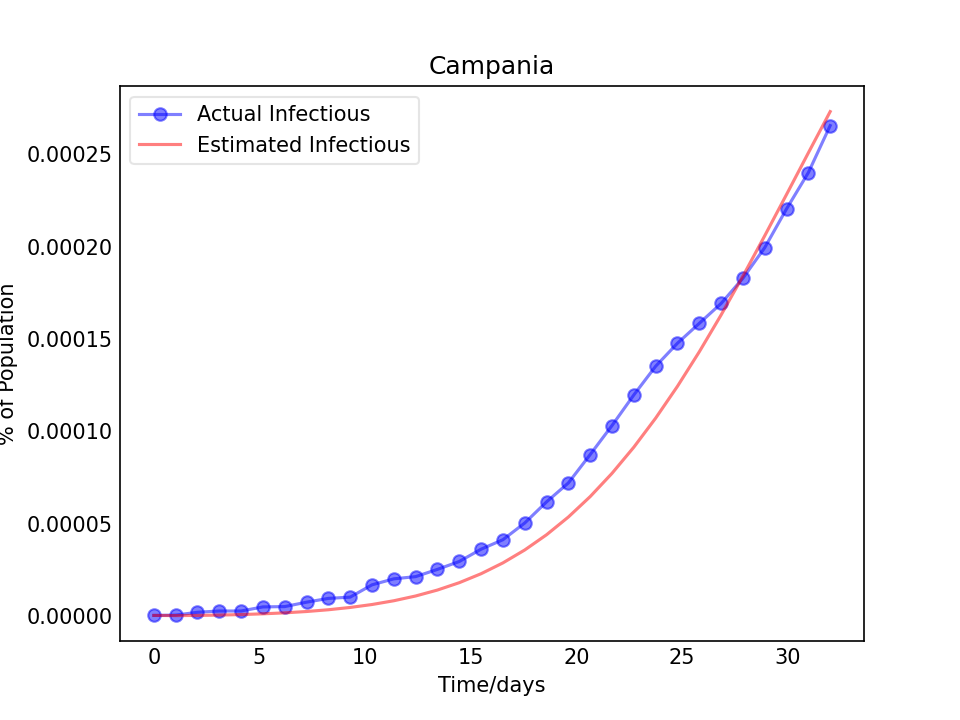}
	\caption{The evolution over time of Covid--19 epidemic in Campania based on the parameter values found for the SEIR--SD model by Differential Evolution.}
	\label{figure-campania-2}
\end{figure}

The use of these values leads to model the diffusion of coronavirus Covid--19 in Campania as it is shown in Fig. \ref{figure-campania-2}.


From the figure we can appreciate that the peak of the infection in Campania is expected to take place around April 8, with a number of infectious subjects equal to about 2.349. Then, this number tends to decrease, and approaches zero around June 3.

\section{Conclusions}

This paper has introduced a variant of the SEIR model for epidemics added with Social Distancing in which the social distancing value is variable over time, and has described how this model can be coupled with Differential Evolution for the individuation of its most suitable parameter values. The resulting mechanism has been applied to model the spreading of the coronavirus Covid--19 in Italy and in two of its most important regions, i.e., Lombardy and Campania. 

Consequently, we have been able to evaluate the evolution over time of the daily number of infectious cases from today until the end of virus spreading. This has allowed us to estimate the day(s) in which this number will be at its highest peak, as well as the day in which the infected cases will become very close to zero.


The above results hold true if the current social distancing is strictly kept in next weeks.  

We hope this study can serve as a useful guideline to Italian Government as well to the Government of any other State in which Covid--19  spreading is occurring.

\bibliographystyle{naturemag-doi}
\bibliography{main}

\end{document}